\documentclass{book}
\usepackage{graphicx}
\usepackage[numbers]{natbib}
\usepackage[english]{babel}
\usepackage{wileyvch}

\author{van Dishoeck \& Visser}
\title{Molecular Photodissociation}

\def\grapprox{_>\atop{^\sim}}
\def\apjl{ApJ}%
\def\apj{ApJ}%
\def\apjs{ApJS}%
\def\aplett{Astrophys.\ Lett.}%
\def\aap{A\&A}%
\def\bain{Bull.\ Astron.\ Inst.\ Netherlands}%
\def\nat{Nature}%
\def\araa{ARA\&A}%

\begin{document}

\setcounter{chapter}{3}
\chapter{Molecular photodissociation}






\section{Introduction}

Photodissociation is the dominant process by which molecules are
removed in any region exposed to intense ultraviolet (UV)
radiation. Such clouds of gas and dust are indicated by astronomers
with the generic title `photon-dominated region', or PDR. Originally,
the term PDR referred mostly to dense molecular clouds close to bright
young stars such as found in the Orion nebula.  There are many other
regions in space, however, in which photodissociation plays a crucial
role in the chemistry: this includes diffuse and translucent
interstellar clouds, high velocity shocks, the surface layers of
protoplanetary disks, and cometary and exoplanetary atmospheres.

In the simplest case, a molecule ABC absorbs a UV photon, which
promotes it into an excited electronic state, and subsequently
dissociates to AB + C. In reality, the situation is much more complex
because there are many electronic states that can be excited, with
only a fraction of the absorptions leading to dissociation. Also,
there are many possible photodissociation products, each of which can
be produced in different electronic, vibrational and rotational states
depending on the wavelength of the incident photons. The rate of
photodissociation depends not only on the cross sections for all of
these processes but also on the intensity and shape of the radiation
field at each position in the cloud.  Thus, an acccurate determination
of the photodissociation rate of even a simple molecule like water
involves many detailed considerations, ranging from its electronic
structure to its dissociation dynamics and to the specifics of the
radiation field to which the molecule is exposed.

In this chapter, each of these steps in determining photodissociation
rates is systematically discussed.  Section 4.2 reviews the basic
processes through which small and large molecules can dissociate
following UV absorption. Techniques for determining absorption and
photodissociation cross sections through theoretical calculations and
laboratory experiments are summarized in Sec.\ 4.3. This section also
mentions new experimental developments that determine branching ratios
of the different products. Various interstellar and circumstellar
radiation fields are summarized in Sec.\ 4.4, including their
attenuation due to dust and self-shielding deeper inside the cloud.
The formulas for computing photodissociation rates are contained in
Sec.\ 4.5, together with references to recent compilations of cross
sections and rates.  As a detailed example, recent developments in our
understanding of the photodissociation of CO and its isotopologs are
presented in Sec.\ 4.6. The chapter ends with a brief review of the
photostability of polycyclic aromatic hydrocarbons (PAHs).  Much of
this chapter follows earlier reviews by \cite{vanD88} and
\cite{vanD06}.


\section{Photodissociation processes}

\subsection{Small molecules}

\begin{figure}[ht]
  \centerline{\hbox{
      \includegraphics[width=0.6\textwidth,angle=0]{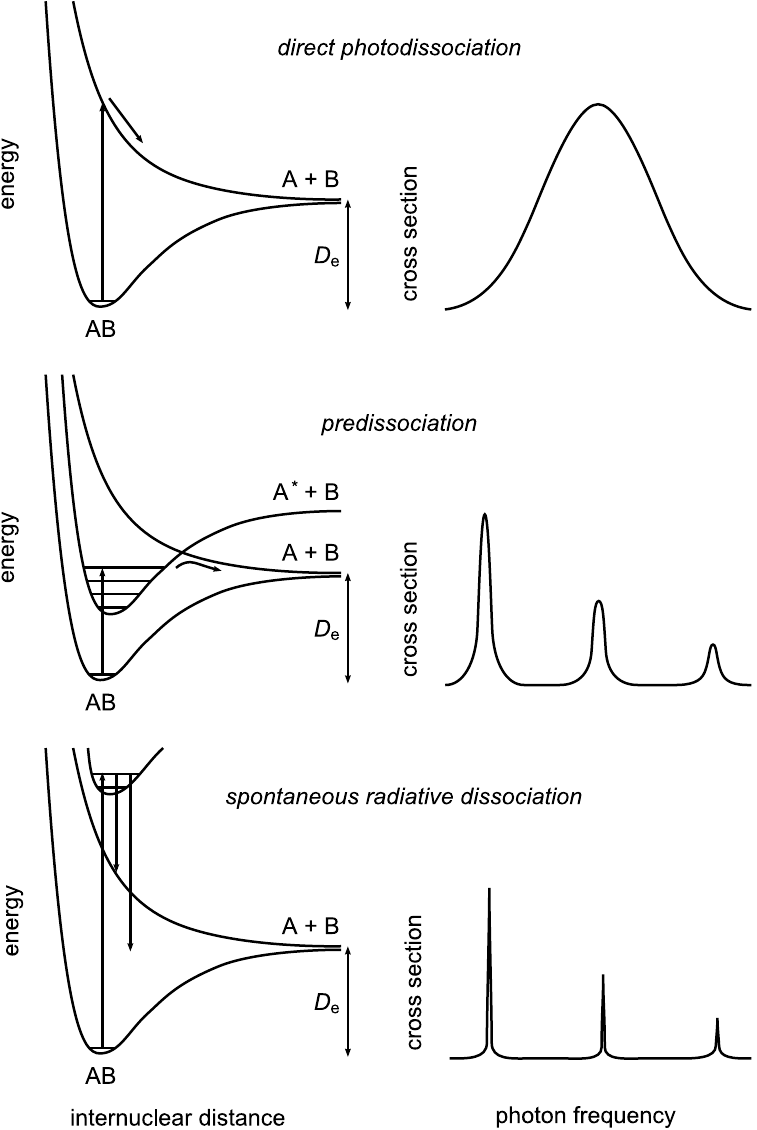}
    }}
  \caption{Photodissociation processes of diatomic molecules and
    corresponding cross sections. From top to bottom: direct
    photodissociation, predissociation and spontaneous radiative
    dissociation (based on \cite{vanD88}).}
  \label{fig:proc}
\end{figure}

The processes by which photodissociation of simple molecules can occur
have been outlined by \cite{Kirby88} and \cite{Schinke93}.  A
summary is presented in Fig.~\ref{fig:proc} for the case of diatomic
molecules. Similar processes can occur for small polyatomic molecules,
especially if the potential surface is dissociative along one
of the coordinates in the multidimensional space.

The simplest process is {\it direct photodissociation}, in which a
molecule absorbs a photon into an excited electronic state that is
repulsive with respect to the nuclear coordinate. Since spontaneous
emission back to the ground state is comparatively slow 
(typical Einstein-A coefficients of $10^9$ s$^{-1}$ compared with
dissociation times of $10^{13}$ s$^{-1}$), virtually all of the
absorptions lead to dissociation of the molecule.  The resulting
photodissociation cross section is continuous as a function of photon
energy, and peaks close to the vertical excitation energy according to
the Franck-Condon principle. The width is determined by the steepness
of the repulsive potential: the steeper the curve, the broader the
cross section. Its shape reflects that of the vibrational wavefunction
of the ground state out of which the absorption occurs. Note that the
cross section is usually very small at the threshold energy
corresponding to the dissociation energy $D_e$ of the molecule; thus,
taking $D_e$ as a proxy for the wavelength range at which dissociation
occurs gives incorrect results.

In the process of {\it predissociation}, the initial absorption occurs
into a bound excited electronic state, which subsequently interacts
non-radiatively with a nearby repulsive electronic state.  An example
of such a type of interaction is spin-orbit coupling between states of
different spin multiplicity. Another example is the non-adiabatic
coupling between two states of the same symmetry. The strength of the
interaction depends sensitively on the type of coupling and on the
energy levels involved, but predissociation rates are typically
comparable to or larger than the rates for spontaneous emission. The
effective photodissociation cross section consists in this case of a
series of discrete peaks, which reflect the product of the oscillator
strength of the initial absorption and the dissociation efficiency of
the level involved. The width is controlled by the sum of the
radiative and predissociation rates and is generally large ($10^{12}$
s$^{-1}$, corresponding to 15 km s$^{-1}$ or more in velocity units).

If the excited bound states are not predissociated, {\it spontaneous
radiative dissociation} can still be effective through emission of
photons into the continuum of a lower-lying repulsive state or the
vibrational continuum of the ground electronic state. The efficiency
of this process is determined by the competition with
spontaneous emission into lower-lying bound states. The
photodissociation cross section again consists of a series of discrete
peaks, but the peaks are not broadened and have widths determined by
the total radiative lifetime (typically $<$0.1 km s$^{-1}$).

\begin{figure}[t]
\centerline{\hbox{
\includegraphics[width=0.6\textwidth,angle=0]{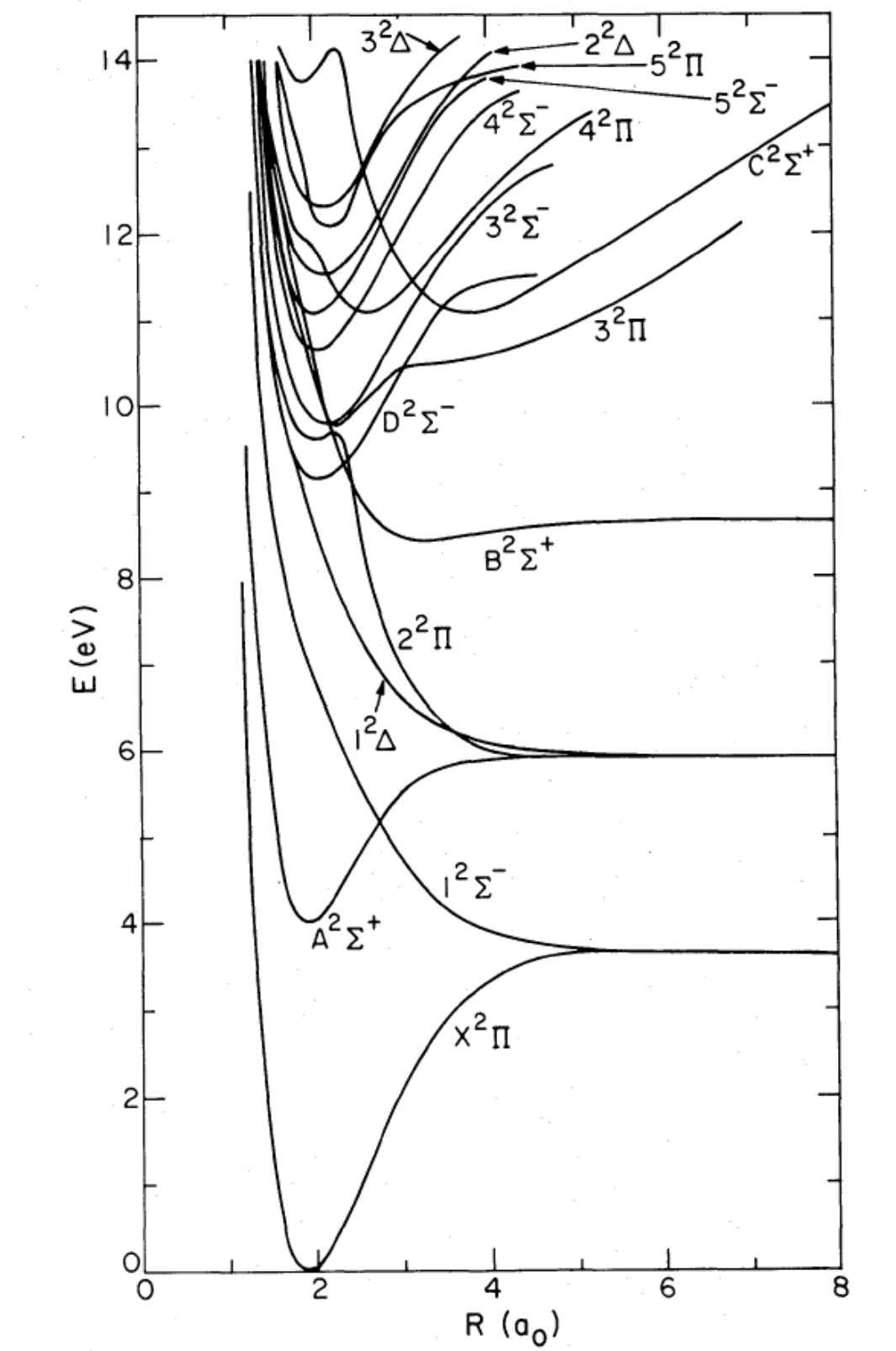}
}}
\caption{Potential energy curves of the OH molecule (reproduced from
  \cite{vanD84}).}
\label{fig:OH}
\end{figure}

Because a molecule has many excited electronic states that can be
populated by the ambient radiation field, in general all of these
processes will occur. As an example, Fig.~\ref{fig:OH} shows the potential
energy curves of the lowest 16 electronic states of the OH radical:
predissociation occurs through the lowest excited $A$~$^2\Sigma^+$
electronic state, whereas direction dissociation can take place
through the 1~$^2\Sigma^-$, 1~$^2\Delta$ and coupled 2 and 3~$^2\Pi$
electronic states. However, usually only one or two of these processes
dominate the photodissociation of a molecule in interstellar
clouds. In the case of OH, these are the direct 1~$^2\Sigma^-$ and
1~$^2\Delta$ channels. However, predissociation through the lower-lying
$A$ state is important in cometary atmospheres and disks around
cool stars, where the radiation field lacks high-energy photons.

The photodissociation of many simple hydrides like H$_2^+$, OH,
H$_2$O, CH, CH$^+$, and NH proceeds mostly through the direct
process. On the other hand, the photodissociation of CO is controlled
by predissociation processes, whereas that of H$_2$ occurs exclusively
by spontaneous radiative dissociation for photon energies below 13.6
eV.  Whether the photodissociation is dominated by continuous or line
processes has important consequences for the radiative transfer
through the cloud (see Sec.\ 4.4.6).

\subsection{Large molecules}

\begin{figure}[t]
\centerline{\hbox{
\includegraphics[width=0.8\textwidth,angle=0]{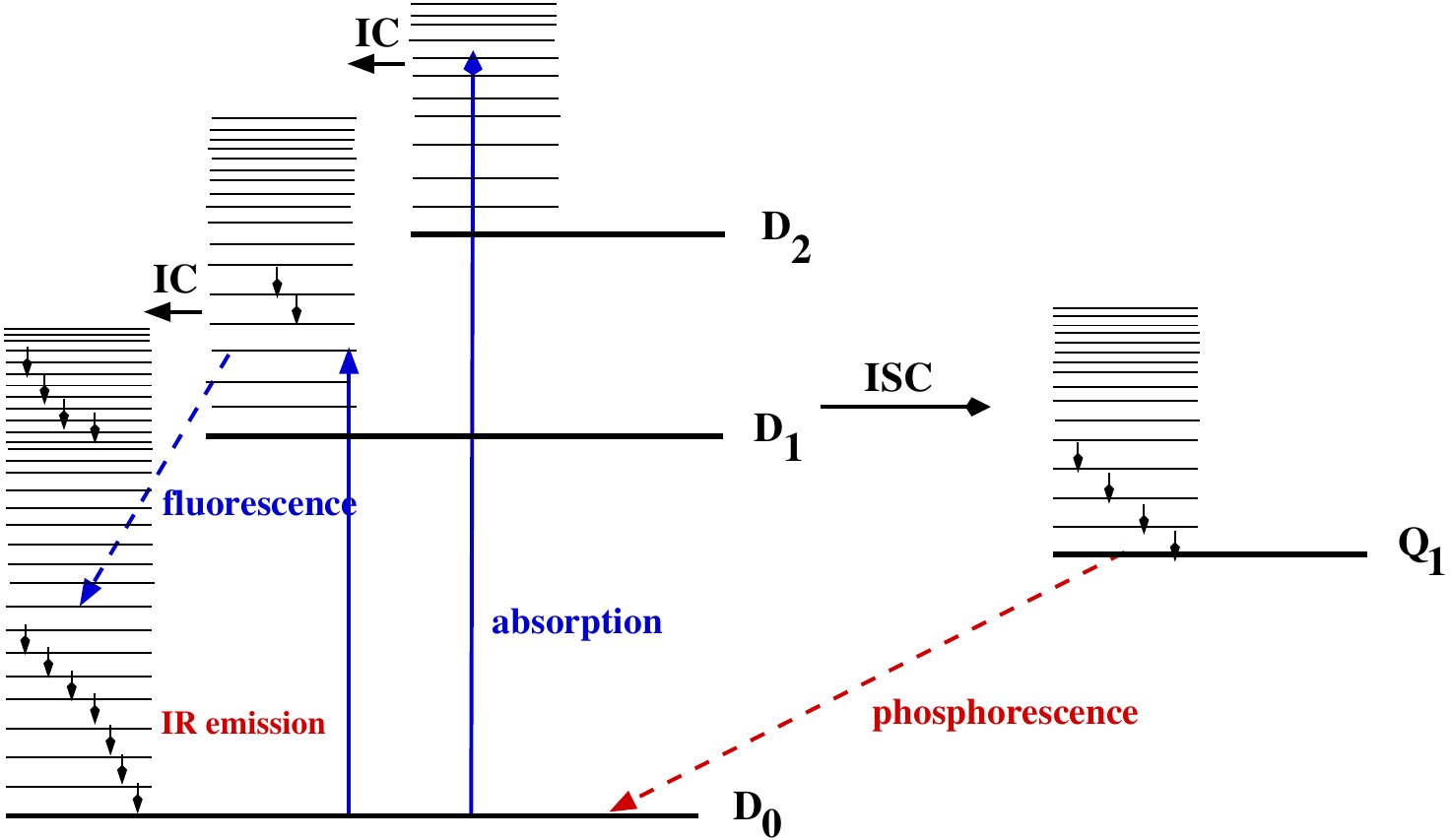}
}}
  \caption{Radiative and radiationless decay of large
    molecules, illustrating the processes of internal conversion (IC),
    intersystem crossing (ISC), fluorescence, phosphorescence and
    infrared emission. This example is for a large ion such as PAH$^+$
    with a ground state with one unpaired electron (doublet spin
    symmetry) denoted as D$_0$, and doublet and quartet excited
    electronic states. Various vibrational levels within the
    electronic states are indicated.}
\label{fig:proclarge}
\end{figure}

For large molecules, the same processes as illustrated in
Fig.~\ref{fig:proc} can occur, but they become less and
less likely as the size of the molecule increases.  This is because
the number of vibrational degrees of freedom of a non-linear
molecule, $3N-6$, increases rapidly with the number of atoms $N$.
Each of these modes has
many associated vibrational levels, with quantum numbers $v$
increasing up to the dissociation energy of the potential. For
sufficiently large $N$, the density of vibrational levels becomes so
large that they form a quasi-continuum with which the excited states
can couple non-radiatively (Fig.~\ref{fig:proclarge}). Through this
so-called process of {\it internal conversion}, the molecule ends up
in a highly excited vibrational level of a lower-lying electronic state. The
probability is small that the molecule will find its way across the
multi-dimensional surface to a specific dissociative mode, so the most
likely outcome is that the molecule ends up in an excited bound level
and subsequently relaxes by emission of infrared
photons. Alternatively, the excited molecule can fluoresce down to the
ground state in a dipole-allowed electronic transition, or it can
undergo so-called intersystem crossing to an electronic state with a different
spin multiplicity from which it can phosphoresce down in a
spin-forbidden transition.

Statistical arguments using (modified) Rice-Ramsperger-Kassel-Marcus
(RRKM) theory suggest that $N$-atom molecules with $N$ $\grapprox$25
are stable with respect to photodissociation, although this number
depends on the structure and types of modes of the molecule involved
\cite{Chowdary08}. Thus, long carbon chains may have a different
photostability than ring molecules. PAHs are a specific class of large
molecules highly relevant for astrochemistry, and are discussed
further in Sec.\ 4.7. Very few gas-phase experiments exist on such large
molecules to test these theories. 

Ashfold et a. \cite{Ashfold10} have recently pointed out that many
intermediate-sized molecules with $N\approx 10$, including complex
organic molecules such as alcohols, ethers, phenols, amines and
N-containing heterocycles, have dissociative excited states that
greatly resemble those found in smaller molecules like H$_2$O and
NH$_3$. Experiments show that H-loss definitely occurs in these larger
systems.

The ionization potentials of large molecules lie typically around 7--8
eV. Most of the absorptions of UV photons with larger energies are
expected to lead to ionization but a small fraction can also result in
dissociation through highly excited neutral states lying above the
ionization threshold or through low-lying dissociative states of the
ion itself. These branching ratios are not well determined
experimentally.

\section{Photodissociation cross sections}

Quantitative information on photodissociation cross sections comes
from theory and experiments.  Theory is well suited for small
molecules, in particular radicals and ions that are not readily
produced in the gas phase.  Laboratory measurements of absorption
spectra over a wide photon range exist primarily for chemically stable
molecules, both small and large. In the following, we first review the
theoretical methods of computing potential energy surfaces and cross
sections, and then discuss the experimental data.

\subsection{Theory}

All information about a molecule can be contained in a wave function
$\Psi$, which is the solution of the time-independent Schr\"odinger
equation
$$ {\cal H} \Psi({x},{R}) = E \Psi ({x},{R}) \eqno(1) $$
Here ${x}$ stands for the spatial and spin coordinates of
the $n$ electrons in the molecule and ${R}$ denotes the positions of all
$N$ nuclei in the molecule.  The total Hamiltonian ${\cal H}$ consists
of the sum of the kinetic energy operators of the nuclei $\alpha$ and
the electrons and of their potential energies due to mutual
interactions. Equation (1) is a $(3n+3N)$-dimensional second order
partial differential equation that cannot be readily solved, even for
small molecules.

\subsubsection{Born-Oppenheimer approximation}

Because the mass of the nuclei is much larger than that of the
electrons, the nuclei move slowly compared with the electrons.  Most
molecular dynamics studies therefore invoke the Born-Oppenheimer
approximation for separating the nuclear and electronic coordinates:
$$ \Psi ({x},{R}) = \Psi^{\rm el}({x};{R}) \Psi^{\rm nuc}({x},{R}) \eqno(2) $$
where the electronic energies $E^{\rm el} ({R})$ (also called the potential
energy curves or surfaces) are determined by solving the electronic
eigenvalue equation with the nuclei held fixed:
$$ {\cal H}^{\rm el} \Psi^{\rm el}({x};{R}) = E^{\rm el}({R}) \Psi^{\rm el}({x};{R}) \eqno(3) $$
Note that $\Psi^{\rm el}$ now only depends parametrically on the nuclear
positions $R$.  Substituting Equation (2) into (1) and using
(3) gives
$$ \bigl [-\sum_\alpha (\frac{1}{2}M_\alpha) \nabla_\alpha^2 + E^{\rm el}(R) - E \bigr ] \Psi^{\rm nuc}(R) = 0 \eqno(4)$$
where the sum is over nuclei $\alpha$ with mass $M_\alpha$.  In this
equation, we use the assumption $\nabla_\alpha^2 \Psi^{\rm el} \Psi^{\rm
  nuc} = \Psi^{\rm el} \nabla_\alpha^2 \Psi^{\rm nuc}$, inherent in the
Born-Oppenheimer approximation. Coupling terms involving
$\nabla_\alpha \Psi^{\rm el}$ are called non-adiabatic terms and their
neglect is usually justified.  Atomic units (a.u., not to be confused
with astronomical units or arbitrary units) have been adopted, which
have $\hbar=m_{\rm e}=e=1$. The unit of distance is then 1 a.u. =
0.52918 \AA\ (also called the Bohr radius or $a_0$) and the unit of
energy is 1 a.u. = 27.21 eV (or Hartree).

Consider for simplicity a diatomic molecule with internuclear distance
$R$. The probability of an electronic transition from state $i$ to
state $f$ is governed by the magnitude of the electronic
transition dipole moment $D(R)$, which can be computed from
$$ D(R) = < \Psi^{\rm el}_f (r;R) \vert d \vert \Psi^{\rm el}_i (r;R) > \eqno(5)$$
where the integration is performed over the electron coordinate
space and $d$ is the electric dipole moment operator in atomic
units. 

The photodissociation cross section (in cm$^2$) following absorption
from a bound vibrational level $v''$ of the ground electronic state
into the vibrational continuum $k'$ of an upper state at a transition
energy $ \Delta E$ is then given by

$$ \sigma_{v''} (\Delta E) = 2.69 \times 10^{-18} g \Delta E \vert
<\Psi^{\rm nuc}_{k'}(R) \vert D(R) \vert \Psi^{\rm nuc}_{v''}(R) > \vert^2 \eqno(6)$$
where the integration is over the nuclear coordinate $R$. Here 
$g$ is a degeneracy
factor (equal to 2 for $\Pi \leftarrow \Sigma $ transitions, 1 otherwise)
and all quantities are in atomic units.  Similarly, the absorption
oscillator strength between two bound states is 
$$f_{v'v''} = {2\over 3} g \Delta E_{v'v''} \vert
<\Psi^{\rm nuc}_{k'}(R) \vert D(R) \vert \Psi^{\rm nuc}_{v''}(R) > \vert^2 \eqno(7)$$
Theoretical calculations of photodissociation cross sections thus
consist of two steps: (1) calculation of the electronic potential
curve or surfaces and (2) solution of the nuclear motion under the influence of
the potentials.

\subsubsection{Electronic energies: method of configuration interaction}

 Step 1 is the calculation of the electronic potential energy surfaces
 $E(R)$ and transition dipole moments $D(R)$ connecting the excited
 states with the ground state as functions of the nuclear coordinate
 $R$. It is important to realize that most aspects of chemistry deal
 with small energy differences between large numbers. For example, the
 binding energy of a molecule or the excitation energy to the first
 excited state is typically only 0.5\% of the total energy of the
 molecule. Thus, the difficulty in quantum chemistry is not only in
 dealing with a $3n+3N$ many-body problem, but also in reaching
 sufficient accuracy in the results. For example, the Hartree-Fock
 method, in which all electrons are treated as independent particles
 by expanding the molecular wave function in Equation (2) in a product of $n$
 one-electron wave functions, fails because it neglects the
 `correlation energy' between the electrons: when two electrons come
 close together, they repel each other. This correlation energy is
 again on the order of a fraction of a percent of the total energy.

Over the last decades, different quantum chemical techniques have been
developed that treat these correlations, and there are several publicly
available programs. Most of the standard packages, however, such as
the popular GAUSSIAN package, or packages
based on density functional theory \cite{teVelde01,Shao06}, are only suitable
and well-tested for the ground electronic state and for closed-shell
electronic structures. More sophisticated techniques based on the
CASSCF (complete active space self-consistent field) or coupled
cluster methods work well for the lowest few excited electronic states
of a molecule, but these states often still lie below the dissociation
energy and do not contribute to photodissociation. Accurate
calculation of the higher-lying potentials through which
photodissociation can proceed requires multi-reference configuration
interaction (MR-CI) techniques \cite{Buenker74,Buenker78}, 
for which only a few packages exist
(e.g., MOLPRO, \cite{Werner10}).

In the CI method, the wave function is expanded into an orthonormal set
of $M$ symmetry-restricted configuration state functions (CSFs).
$$ \Psi^{\rm el}({ x_1} ... { x_n}) = \sum_1^M c_s \Phi_s({ x}_1 ... {x}_n) \eqno(8) $$
The CSFs or `configurations' are generally linear combinations of
Slater determinants, each combination having the symmetry and
multiplicity of the state under consideration. The Slater determinants
are constructed from an orthonormal set of one-electron molecular
orbitals (MOs; obtained from an SCF solution to the Hartree-Fock
equations), which in turn are expanded in an elementary set of atomic
orbitals (AOs; called `the basis set') centered on the atomic
nuclei. The larger $M$ is, the more accurate the results.

In the multi-reference technique, a set of configurations is chosen to
provide the reference space. For example, the $X$~$^2\Pi$ ground state
of OH has the main configuration $1\sigma^2 2\sigma^2 3\sigma^2
1\pi^3$, where $\sigma$ and $\pi$ are the molecular orbitals. The
reference space could be chosen to consist of all possible
configurations of the same overall symmetry that have a coefficient
$c_s $ greater than some threshold in the final CI wave function. For
the OH example, this includes configurations such as $1\sigma^2
2\sigma^2 3\sigma^1 4\sigma^1 1\pi^3$ in which one electron is excited
from the highest occupied MO (HOMO) to the lowest unoccupied MO
(LUMO). Alternatively, an `active space' of orbitals can be
designated, e.g., the $2\sigma-4\sigma$ and $1\pi-2\pi$ MOs in
the case of OH, within which all configurations of a particular
symmetry are considered. The CI method then generates all single and
double (and sometimes even higher) excitations with respect to this
set of reference configurations, several hundreds of thousands of
configurations in total, and diagonalizes the corresponding
matrix. The resulting eigenvalues are the different electronic states
of the molecule, with the lowest energy root corresponding to the
ground state.

The quality of an MR-CI calculation is ultimately determined by the
choice of the atomic orbital basis set, the choice of the reference
space of configurations, and the number of configurations included in
the final CI. The basis set needs to be at least of `triple-$\zeta$'
quality (i.e., each occupied atomic orbital $1s, 2p, ...$ is represented
by three functions with optimized exponents $\zeta$). Also,
polarization and Rydberg functions (i.e., functions which have higher
quantum numbers than the occupied orbitals, e.g., $2p$ for H, $3d$ and $4s$
for C) need to be added. The number of AOs chosen in the basis set
determines the number of MOs, which in turn determines the number of
configurations. For a typical high-quality AO basis set, the latter
number is so large, of order $10^8$, that some selection of
configurations needs to be made in order to handle them with a
computer.  For example, all configurations which lower the energy by
more than a threshold value (in an eigenvalue equation with the
reference set) can be chosen to be included in the final CI matrix.

Once the electronic wave functions $\Psi^{\rm el}$ for the ground and
excited electronic states have been obtained, the expectation values
of other operators, such as the electric dipole moment or the
spin-orbit coupling, can be readily calculated.  


\subsubsection{Nuclear dynamics: oscillator strengths and cross sections}

Step 2 in calculating photodissociation cross sections
consists of solving
Eq.\ (4) to determine the nuclear wave functions using 
the electronic potential curves from Step 1. Often
the nuclear wave function is first separated into a radial and an
angular part. If the angular part is treated as a rigid rotor, the
radial part can be solved exactly by one-dimensional numerical
integration for a diatomic molecule so that no further uncertainties
are introduced. Once the wave functions for ground and excited states
are obtained, the cross sections can be computed
according to Eq.\ (6).

For indirect photodissociation processes, the oscillator strengths
into the discrete upper levels are computed according to Eq.\ (7)
for each vibrational level $v'$.  If the coupling between the upper
bound level with the final dissociative continuum (as computed in Step
1) is weak, first order perturbation theory can be used to calculate
the predissociation rates $k^{\rm pr}$ in s$^{-1}$. The
predissociation probability $\eta_u$ of upper level $u$ is then
obtained by comparing $k^{\rm pr}$ with the inverse radiative lifetime
of the molecule $A^{\rm rad}$: $\eta_u=k^{\rm pr}/(k^{\rm pr} + A^{\rm
  rad})$. In our example of OH, this approximation works well for the
calculation of the predissociation of the $A$~$^2\Sigma^+$ state for
$v'\geq 2$.  If the coupling is strong, as is the case for the OH 2
and 3~$^2\Pi$ states, the coupled equations for the excited states
(i.e., going beyond the Born-Oppenheimer approximation) have to be
solved in order to compute the cross sections.

For (light) triatomic molecules, the calculation of the full 3D
potential surfaces and the subsequent dynamics on those surfaces is
still feasible, albeit with significantly more effort involved
\cite{Kroes93,Schinke93,Bearda95}. For multi-dimensional
potential surfaces, often time-dependent wave packet propagation
methods are preferred to solve the dynamics rather than the
time-independent approach.  Besides accurate overall cross sections,
such studies give detailed insight into the photodissociation
dynamics. For example, the structure seen in the
photodissociation of H$_2$O through the first excited $\tilde A$ state
is found to be due to the symmetric stretch of the excited molecule
just prior to dissociation.

\begin{figure}[t]
\centerline{\hbox{
\includegraphics[width=0.6\textwidth,angle=-90]{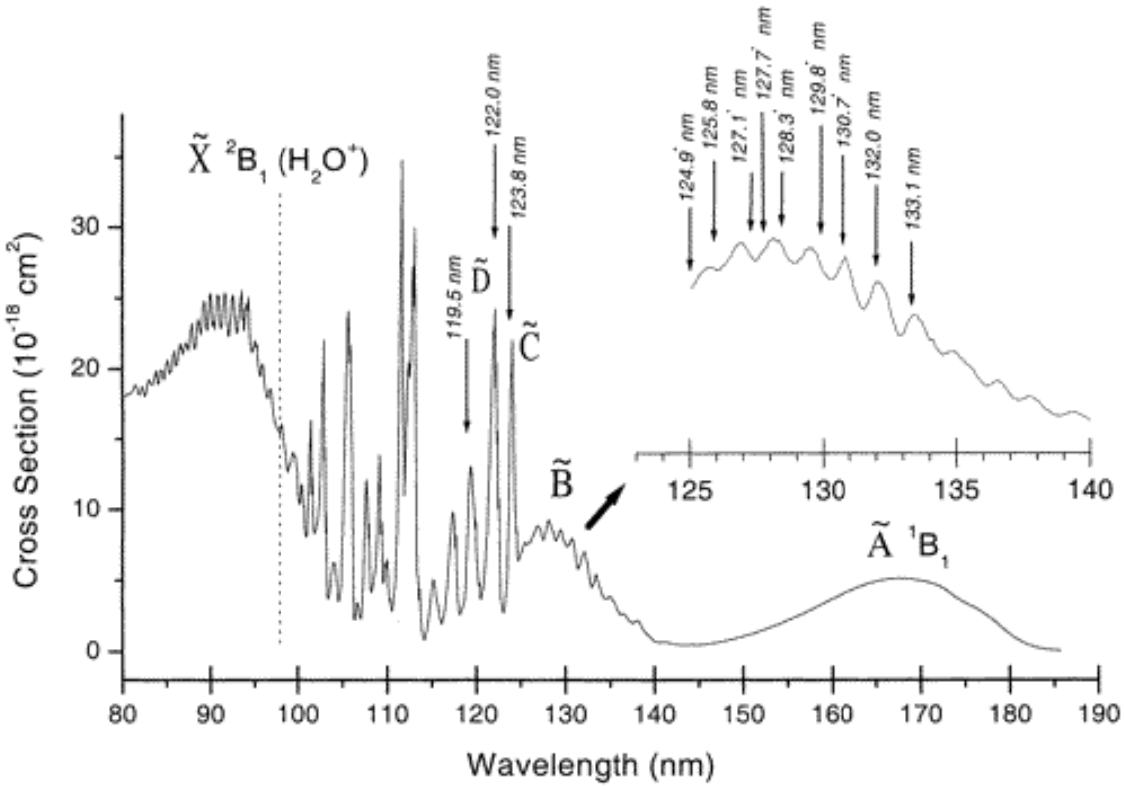}
}}
\caption{Absorption spectrum of gaseous H$_2$O illustrating the
  different electronic states through which photodissociation can
  occur (reproduced from \cite{Fillion01}).}
\label{fig:H2O}
\end{figure}

For larger polyatomic molecules, such fully flexible calculations are
no longer possible and one or more nuclear coordinates need to be
frozen.  In practice, often the entire nuclear dynamics calculations
are skipped and only electronic energies and oscillator strengths at
the ground state equilibrium geometry are computed. Assuming that all
absorptions into excited states with vertical excitation energies
larger than $D_e$ lead to dissociation (i.e., $\eta_u=1$), this method
provides a rough estimate (upper limit) on the dissociation rate
\cite{vanHemert08}. This method works because of the Franck-Condon
principle, which states that the highest cross sections occur when the
excitation energies are vertical, that is, nuclear coordinates do not
change between lower and upper states (Fig.~\ref{fig:proc}). Because
only a single nuclear geometry has to be considered, the amount of
work is orders of magnitude less than that of a full electronic and
nuclear calculation.

The overall accuracy of the cross sections and oscillator strengths is
typically 20--30\% for small molecules in which the number of
active electrons is at most $\sim$30. On the order of 5--10 electronic states
per molecular symmetry can be computed with reasonable
accuracy. Transition energies to the lowest excited states are
accurate to $\sim$0.1 eV, those to higher states to 0.2--0.3 eV. Thus,
quantum chemistry cannot predict whether a molecule has an electronic
state whose excitation energy matches exactly that of, say, Lyman
$\alpha$ at 10.2 eV. However, if it has an electronic state with a
vertical energy close to 10 eV, and absorption into the state is
continuous, theory can firmly state that there is a significant
cross section at Lyman $\alpha$. An example is the 1~$^2\Delta$ state
of OH: even with $\pm$30 \AA\ uncertainty in position, the cross
section at 1215.6 \AA\ is found to lie between 1 and $2\times
10^{-18}$ cm$^2$.

Virtually all molecules have more electronic states below 13.6 eV than
can be computed accurately with quantum chemistry. Take our OH
example: there is an infinite number of Rydberg states converging to
the ionization potential at 13.0 eV. However, because the intensity of the
radiation field usually decreases at shorter wavelengths, these higher
states do not significantly affect the overall photodissociation
rates. Their combined effect can often be taken into account
through a single state with an oscillator strength of 0.1 lying around
the ionization potential. 


\subsection{Experiments}

Laboratory measurements of absorption cross sections have been
performed for many chemically stable species, including
astrophysically relevant molecules such as H$_2$O, CO$_2$, NH$_3$, and
CH$_4$. Most of these experiments have been performed at rather low
spectral resolution, where the individual ro-vibrational lines are not
resolved. Figure~\ref{fig:H2O} shows an example of measurements for H$_2$O.  A
broad absorption continuum is observed between 1900 and 1200 \AA, with
discrete features superposed at shorter wavelengths.  Note that the
electronic states responsible for the absorptions at the shortest
wavelengths have often not yet been identified spectroscopically.

The absorption of a photon can result in re-emission of another
photon, dissociation, or ionization of the molecule, and most
experiments do not distinguish between these processes. If the photon
energy is below the first ionization potential and if the absorption
is continuous, photodissociation is likely to be the dominant process.  If
not, additional information is needed to infer the dissociation
probabilities $\eta_u$. For the H$_2$O example, the broad
continuum at 1900--1400 \AA\ corresponds to absorption into the
$\tilde A$~$^1B_1$ state, which is fully dissociative.  However, this is
not necessarily true for the higher-lying discrete absorptions. An
example is provided by the case of NO for which fluorescence cross
sections have been measured directly and found to vary significantly
from band to band, with $\eta_u$ significantly less than unity for
some bands \cite{Lee84}.

Experiments typically quote error bars of about 20\% in their cross
sections if the absorption is continuous. For discrete absorptions,
high spectral resolution and low pressures are essential to obtain
reliable cross sections or oscillator strengths, since saturation
effects can easily cause orders of magnitude errors.

Most broad-band absorption spectra date from the 1950s-1980s and have
been summarized in various papers and books
\cite{Lee84,Hudson71,Okabe78}. An electronic compilation is provided
through the MPI-Mainz UV-VIS spectral atlas. Since about 1990,
emphasis has shifted to experiments at single well-defined wavelengths
using lasers, in particular at 1930 and 1570 \AA. The aim of these
experiments is usually to study the dynamics of the photodissociation
process at that wavelength, rather than obtaining cross sections. Such
studies typically target the lower excited electronic states since
lasers or light sources at wavelengths $<$1200 \AA\ are not commonly
available except in specialized laboratories \cite{Ubachs94}.  While
these sophisticated experiments have provided much basic insight into
photodissociation dynamics, the restriction to only one or two
wavelengths and lack of cross section data makes them of limited value
for interstellar applications.

With the recent advent of powerful synchrotron light sources providing
continuous radiation over a large wavelength range, the pendulum is
swinging back to more astrophysically relevant results. For example,
photodissociation cross sections of molecules such as CH$_3$OH are
being measured in the X-ray regime up to 100 eV
\cite{Pilling07}. High-resolution data in the FUV regime up
to 13.6 eV are eagerly awaited.

\subsection{Photodissociation products}

Inclusion of photodissociation reactions in chemical networks requires
knowledge not only of the rate of removal of species ABC, but also of
the branching ratios to the various products, AB + C, A + BC or AC +
B. Each of these species can be produced with different amounts of
electronic, vibrational and/or rotational energy, with the absolute and
relative amounts depending on wavelength. None of this detail is
captured in the standard tabulations, which, at best, give branching
ratios between products integrated over all wavelengths or at a single
wavelength, and neglect any vibrational or rotational excitation.

A great number of very elegant modern experiments are now available to
explore the fragments, based on a range of photofragment translational
spectroscopy and time-resolved photon-electron spectroscopy methods
\cite{Ashfold10}.  Product imaging techniques
date back to the 1980s and characterize both the velocities,
internal energies and angular distributions of the photodissociation
products \cite{Chandler87}.
  
\begin{figure}[ht]
\centerline{\hbox{
\includegraphics[width=0.5\textwidth,angle=0]{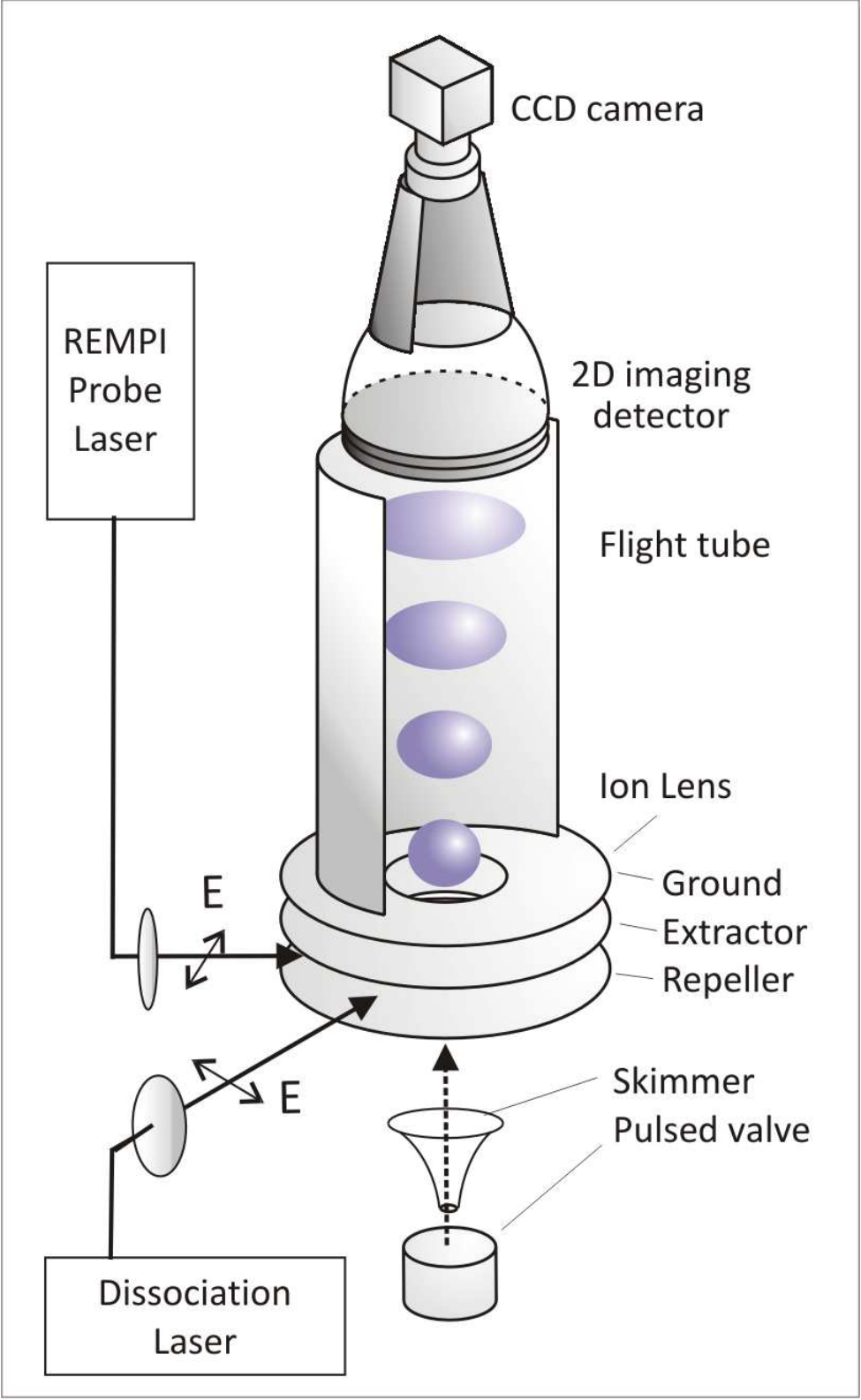}
}}
\caption{Experimental setup for velocity map imaging of photodissociation
products (based on \cite{Eppink97}).}
\label{fig:Parker}
\end{figure}

Velocity map imaging is a recent example of such a technique well suited
for probing photodissociation (Fig.~\ref{fig:Parker})
\cite{Eppink97,Ashfold06}.  In this method, 
a molecular beam containing
the molecule of interest is created by a pulsed supersonic expansion,
which ensures internal state cooling.  After passing through a skimmer
and a small hole in the repeller plate lens of an ion lens assembly
the beam is crossed by a dissociation laser beam to form neutral
fragments, which are immediately state-selectively ionized by a probe
laser beam using resonance enhanced multi-photon ionization (REMPI).
The ions retain the velocity information of the nascent
photofragments.  Velocity map imaging uses a special configuration of
an electrostatic immersion ion lens (combination of repeller,
extractor, and ground electrodes) to ensure mapping of the ion
velocity independent of its position of formation.  After acceleration
through the ion lens, the particles are mass-selected by time of flight
upon reaching the surface of a two-dimensional imaging detector,
which converts ion impacts to light flashes recorded by a CCD camera.
The two-dimensional image can then be converted to its
desired 3-D counterpart by using an inverse Abel transformation.  The
lens effectively reduces the `blurring' of the images.


Another variant on this technique is Rydberg H-atom time of flight
spectroscopy. Here the nascent H atoms produced by photodissociation
in the ground $n=1$ state are excited into the $n=2$ state using a
Lyman $\alpha$ laser at 1216 \AA\ and subsequently to a high Rydberg
state with $n=30$--80 using longer wavelength (e.g., 3650 \AA)
radiation. The neutral Rydberg H atoms then reach a microchannel
plate where they are field-ionized and detected.

A completely different set of experiments has been applied to study
the branching ratios of large carbon chain molecules. Here, high
velocity collisions combined with an inverse kinematics scheme have
shown that the most favorable channels are H production from C$_n$H
and C$_3$ production from C$_n$ \cite{Chabot10}. Otherwise the
fragmentation behavior is largely statistical.

An example of the complexity of the situation is provided by the OH
and H$_2$O cases. For OH, the potential curves in Fig.~\ref{fig:OH}
show that absorption into the 1~$^2\Sigma^-$ state results in
ground-state O($^3$P) + H, but absorption into the higher lying
$^2\Delta$, $^2\Pi$ and $^2\Sigma^+$ states in excited O($^1$D) and
O($^1$S). The cross sections into these states are such that about
equal amounts of O($^3$P) and O($^1$D) are produced in the general
interstellar radiation field, but only 5\% of O($^1$S)
\cite{vanD84}. An astrophysical confirmation of this prediction is
the detection of the red line of atomic O ($^1$D $\to$ $^3$P) at 6300
\AA\ from protoplanetary disks \cite{Storzer98,Acke06}. On the other
hand, for radiation fields dominated by lower photon energies such as
in comets exposed to solar radiation, the main product is
ground-state O($^3$P).

H$_2$O is one of the few polyatomic molecules for which the product
distribution has been well characterized as a function of energy. Both
theory and experiment have shown that absorption at 6--8 eV into the
first $\tilde A$ electronic state produces OH in the ground
electronic $^2\Pi$ state with some vibrational excitation but little
rotational excitation. In contrast, absorption at 9--11 eV into the
second $\tilde B$ electronic state produces OH in highly excited
rotational levels of the $A ^2\Sigma^+$ excited electronic state, but with low
vibrational excitation. Absorption into even higher excited states results
in O + H$_2$ or O + H + H products, rather than OH + H.
For other polyatomic molecules, experimental information is spotty at best.
Early experiments were often performed at high pressures, where the
observed products could result from subsequent chemical reactions in
the gas that would not occur at the low densities in space \cite{Okabe78}.  

Some guess of the most likely products can be made on the basis of
product energies (when does a product channel open up?) and
correlation diagrams \cite{Ashfold10}, but these
techniques are usually limited to the lowest-lying channels that may
not dominate the overall dissociation. Often, the astrochemical
databases simply assume a statistical distribution over the various
products averaged over all photon energies.
Note that the information on the product distribution of species like
CH$_3$OH is relevant not only for gas-phase chemistry but even more so
for ice chemistry \cite{Oberg09}.


\section{Astrophysical radiation fields}

\subsection{General interstellar radiation field}

Determinations of the average intensity of the interstellar radiation
field (ISRF) fall into two main categories (see \cite{Black94} and
\cite{vanD94} for critical reviews). The first method
is to estimate the number and distribution of hot stars (O, B, A,
... type) in the Galaxy, use a model for the dust distribution and
properties to determine the absorption and scattering of the stellar
radiation, and then sum their fluxes to determine the energy density
at a typical interstellar location. This method dates back to
\cite{Habing68} and \cite{Draine78}, and has most recently been revisited
by \cite{Parravano03}. The stellar fluxes used in these models are a
combination of observed fluxes of early-type stars and extrapolations
to shorter wavelengths using model atmospheres.

\begin{figure}
\centerline{\hbox{
\includegraphics[width=0.8\textwidth]{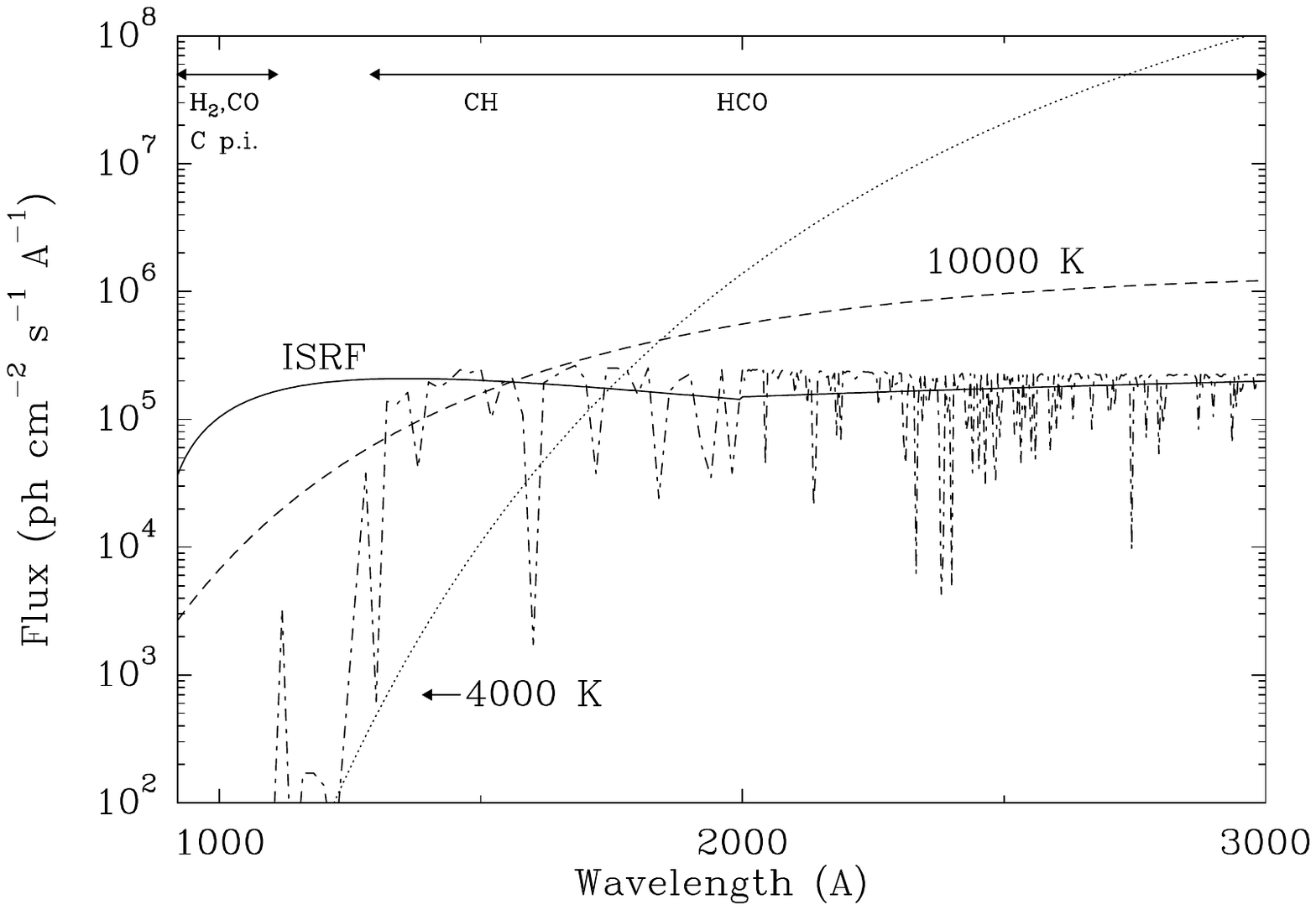}
}}
\caption{Comparison of the general interstellar radiation field of
  Draine (1978) (extended for $\lambda > 2000$ \AA\ using the
  representation of \cite{vDB82}) with various stellar radiation
  fields scaled to have the same integrated intensity from 912--2050
  \AA.  The scaled NEXTGEN model radiation field of a B9.5 star
  \cite{Hauschildt99} is included as well (dash-dotted).  The
  wavelength ranges where the photodissociation of some important
  interstellar molecules occurs are indicated (reproduced from
  \cite{vanD06}).}
\label{fig:radfield}
\end{figure}

The second method is to use direct measurements of the UV radiation
from the sky at specific wavelengths \cite{Henry80}. Also in this
method, model stellar atmospheres are used to provide information at
wavelengths not directly observed. In both methods, direct starlight
from early B stars is found to dominate the measured interstellar
flux.  Specifically, most of the flux at a given point in space comes from the
sum of discrete sources within 500 pc distance from that point.

The various estimates of the local ISRF in the solar neighborhood
agree remarkably well, within a factor of two.  The median value found
by \cite{Parravano03} is within 10\% of the mean energy density of
$U=8\times 10^{-17}$ erg cm$^{-3}$ \AA$^{-1}$ of \cite{Draine78},
averaged over the 912--2070 \AA\ FUV band ($\Delta_{\rm FUV}$). The
strength of the one-dimensional interstellar radiation field is often
indicated by a scaling factor $G_0$ with respect to the flux $F$ in
the Habing field ($F_{\rm FUV}=cU_{\rm FUV}\Delta_{\rm FUV}$) of
$1.6\times 10^{-3}$ erg cm$^{-2}$ s$^{-1}$, which is a factor of 1.7
below that of \cite{Draine78}.  Thus, the standard Draine field
implies $G_0=1.7$.  Care should be taken that photodissociation rates
used in astrochemical models refer to the same radiation field.

If a cloud is located close to a young, hot star, the intensity
incident on the cloud boundary may be enhanced by orders of magnitude
compared with the average ISRF. A well-known example is the Orion Bar,
where the intensity is enhanced by a factor of $4.4\times 10^4$ with
respect to the Draine field. In contrast, at high latitudes where few
early-type stars are located, the intensity may be factors of 5--10
lower in the FUV band than that of the standard ISRF field. Throughout
the galactic plane and over time scales of a few Gyr, variations in
the local energy density by factors of 2--3 are expected due to the
birth and demise of associations containing high-mass O and B stars
within about 30 Myr. Moreover, the ratio of the highest-energy photons
capable of dissociating H$_2$ and CO and those in the more general FUV
range may vary by a factor of 2 in time and place \cite{Parravano03}.


\subsection{Stellar radiation fields}

The surface layers of disks around young stars are another example of
gas in which the chemistry is controlled by photodissociation. The
illuminating stars range from late B- and early A-type stars (the
so-called Herbig Ae/Be stars) to K- and M-type stars (the T Tauri
stars).  For the latter, if the stellar atmosphere dominates the flux,
orders of magnitude fewer high-energy photons are available to dissociate
the molecules than in the ISRF scaled to the same integrated
intensity (Fig.~\ref{fig:radfield}). In particular, the
number of photons capable of dissociating H$_2$ and CO and ionizing
carbon is greatly reduced \cite{vanD06,Kamp01}. However, if
accretion onto the star is still taking place, the hot gas in the accreting
column produces high-energy photons, including Lyman $\alpha$, which
can dominate the FUV flux \cite{Bergin03}.

\subsection{Lyman $\alpha$ radiation}

Fast shocks (velocities $>$50 km s$^{-1}$) produce intense Lyman
$\alpha$ radiation due to collisional excitation of atomic hydrogen in
the hot gas or recombination of ionized hydrogen
\cite{Hollenbach79,Neufeld89}. Another prominent line in shocks is
the C III resonance line at 977 \AA. Fast shocks can be caused by
supernovae expanding into the interstellar medium, or by fast jets
from protostars interacting with the surrounding cloud. Accretion
shocks onto the young star mentioned above are another example, as are
exoplanetary atmospheres.  Some molecules, in particular H$_2$, CO and
N$_2$, cannot be dissociated by Lyman $\alpha$. Other simple molecules
such as OH, H$_2$O and HCN can.  A recent summary of cross sections at
Lyman $\alpha$ is given by \cite{vanD06}.

\subsection{Cosmic-ray induced photons}

\begin{figure}
\centerline{\hbox{
\includegraphics[width=0.9\textwidth,angle=0]{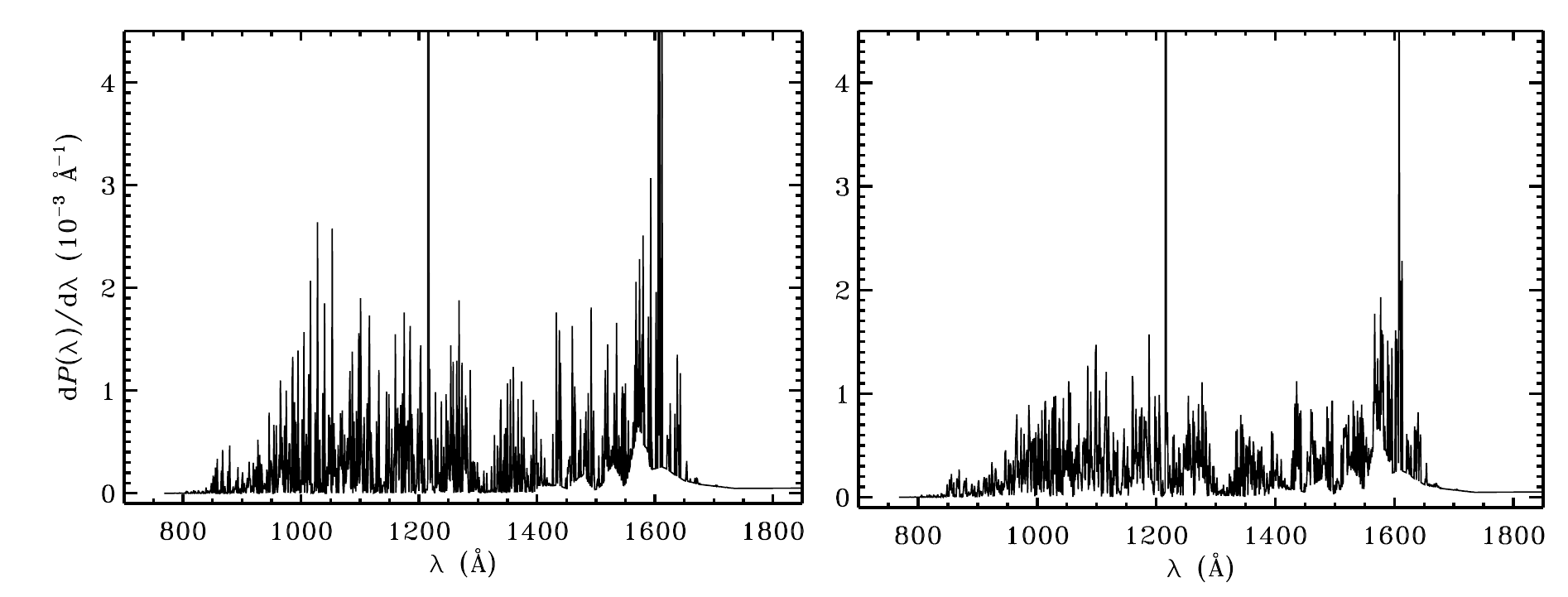}
}}
\caption{Cosmic-ray induced spectrum of H$_2$ assuming that all H$_2$
  is in $J$=0 (left) or distributed over $J$=1 and $J$=0 in a 3:1
  ratio (left). The spectrum shows the number of photons per 1 \AA \ bin for
  initial 30 eV electrons and are normalized to the total number of
  ionizations (figure by Gredel, updated from \cite{Gredel87,Gredel89}).}
\label{fig:CRfield}
\end{figure}

A dilute flux of UV photons can be maintained deep inside
dense clouds through the interaction of cosmic rays with hydrogen
in the following process:
$$ {\rm H\ or\  H_2 + CR} \to {\rm H^+\ or\ H_2^+ + e^*} \eqno(9) $$
$$ {\rm e^* + H_2} \to { \rm H_2^* + e} $$
$$ {\rm H_2^*} \to {\rm H_2 + h\nu} $$
The energetic electrons e$^*$ produced by the cosmic rays excite H$_2$
into the $B$~$^1\Sigma_u^+$ and $C$~$^1\Pi_u$ electronic states, so
that the FUV emission is dominated by Lyman- and Werner-band UV
photons. Higher-lying electronic states contribute at the shortest
wavelengths. As a result, the UV spectrum produced inside dense clouds
resembles strongly that of a standard hydrogen lamp in the
laboratory. Figure~\ref{fig:CRfield} shows the UV spectrum computed
following \cite{Gredel87,Gredel89} for e$^*$ at 30 eV and either all
H$_2$ in $J$=0 or distributed over $J$=1 and $J$=0 in a 3:1
ratio. Note that the precise spectrum does depend on the H$_2$ level
populations and the ortho/para ratio. Because of the highly structured FUV field
and the lack of high-resolution cross sections for most species, the
resulting photodissociation rates may be quite uncertain. On the other
hand, the large number of lines has a mitigating effect.

\subsection{Dust attenuation}

The depth to which ultraviolet photons can penetrate in the cloud
and affect the chemistry depends on (i) the 
amount of photons at the boundary of the cloud; (ii) the
scattering properties of the grains as functions of wavelength;
(iii) the competition between the atoms and molecules
with the grains for the available photons, which, in turn, depends
on the density of the cloud; and (iv) possible self-shielding of the
molecules.

The absorption and scattering parameters of the grains determine the
penetration depths of photons.  The ultraviolet extinction curve shows
substantial variations from place to place, especially at the shortest
wavelengths in the 912--1100 \AA\ region, which can greatly affect
the CO and CN photodissociation rates \cite{Cardelli88,vDB89}.
Similar variations are expected to occur for the albedo $\varpi$ and
the asymmetry parameter $g$ as functions of wavelength, but little is
still known about these parameters at the shortest wavelengths
\cite{Roberge91}.

There is considerable observational evidence that dust grains in the
surface layers of protoplanetary disks have grown from the typical
interstellar size of 0.1 $\mu$m to at least a few $\mu$m in size or
more. These large grains extinguish the UV radiation much less,
allowing photodissociation to penetrate deeper into the disk
\cite{vanD06,Vasyunin11}.

\subsection{Self-shielding}

Because the physical and chemical structure of cloud edges is
controlled by the photodissociation of H$_2$ and CO and
photoionization of C, it is important to pay particular attention to
the details for these molecules. CO is discussed in Sec.\ 4.6. The
photodissociation of H$_2$ occurs by absorptions into the Lyman and
Werner bands at 912--1100 \AA\ followed by spontaneous emission back
to the vibrational continuum of the ground state 
\cite[see also Fig.~\ref{fig:proc}]{Dalgarno70}.  The absorption lines
have typical
oscillator strengths $f\approx0.0001$--0.03, and on average about 10\%
of the absorptions lead to dissociation.  The strongest lines become
optically thick at H$_2$ column densities of $10^{14}$--$10^{15}$
cm$^{-2}$ for a Doppler broadening parameter $b$ of 3 km s$^{-1}$,
implying that molecules lying deeper within the cloud are shielded from
dissociating radiation because all relevant photons have been absorbed at the
cloud edge (`self-shielding').  The photodissociation rate of H$_2$ is
about $4.5\times 10^{-11}$ s$^{-1}$ in the unattenuated interstellar
radiation field, corresponding to a lifetime of about 700 yr.  Inside
the cloud, the lifetime becomes longer by several orders of magnitude
because of the self-shielding process.  The remaing 90\% of the
absorptions are followed by UV fluorescence back into the
bound vibrational levels of the H$_2$ ground state.  

Photoionization of atomic carbon has a continuous cross section of
about $10^{-17}$ cm$^2$ over the 912--1100 \AA\ region. Thus, for
column densities $N({\rm C})>10^{17}$ cm$^{-2}$, carbon starts to
self-shield \cite{Werner70}. Moreover, the saturated absorption
bands of H and H$_2$ over the same wavelength range remove a
considerable fraction of the ionizing radiation (`mutual shielding').

\section{Photodissociation rates}

The photodissociation rate $k_{pd}$ of a molecule by 
continuous absorption is given by
$$ k_{\rm pd}^{\rm cont} = \int \sigma (\lambda) I(\lambda) d \lambda
   \ \ {\rm s^{-1}}, \eqno (10) $$
where $\sigma$ is the cross section for photodissociation in cm$^2$
   and $I$ is the mean intensity of the radiation in photons cm$^{-2}$
   s$^{-1}$ \AA$^{-1}$\ as a function of wavelength $\lambda$ in \AA \
   (Fig.~\ref{fig:radfield}). For the indirect processes of
   predissociation and spontaneous radiative dissociation, the rate of
   dissociation by absorption into a specific level of a bound upper
   state $u$ from lower level $\ell$ is
$$ k_{\rm pd}^{\rm line} = { {\pi e^2} \over {m c^2}} \lambda^2_{u\ell}
        f_{u\ell} \eta_{u} x_{\ell} I(\lambda_{u\ell}) \ \
         {\rm s^{-1}}, \eqno(11) $$
where $f_{u\ell}$ is the oscillator strength, $\eta_u$ is the
dissociation efficiency of the upper level (between 0 and 1),
and $x_{\ell}$ is the fractional population in level $\ell$.

The determination of the cross sections and oscillator strengths from
theory and experiments has been discussed in Sec.\ 4.3. Cross section
databases are available
from \cite{vanD06} at {\tt www.strw.leidenuniv.nl/$\sim$ewine/photo}
and from \cite{Huebner92} at {\tt amop.space.swri.edu} for
cometary species. For more complex species also of atmospheric interest, a
large compilation is available through the MPI-Mainz UV-VIS spectral
atlas at {\tt www.atmosphere.mpg.de/enid/2295} and
{\tt www.science-softcon.de/spectra}.

Photodissociation rates as functions of depth into a cloud using the
standard interstellar radiation field \cite{Draine78} have been
presented for a variety of molecules \cite{vanD88,Roberge91}, and most
recently by \cite{vanD06}. The latter paper and associated website
also summarize rates in cooler radiation fields such as appropriate
for late-type stars. The depth dependence due to continuum extinction
can be represented by a simple exponential function, with exponents
that vary with grain scattering properties and that depend
parametrically on cloud thickness.  A list of cosmic-ray induced rates
for use in dense cloud models has been given by \cite{Gredel89}.

\section{Photodissociation of CO and its isotopologs}

CO is the most commonly observed molecule in interstellar space and
used as a tracer of molecular gas throughout the universe, from local
diffuse clouds to dense gas at the highest redshifts.  Additional
impetus for a good understanding of its photodissociation processes
comes from recent interpretations of oxygen isotope fractionation in
primitive meteorites, which are thought to originate from isotope
selective photodissociation of CO in the upper layers of the disk out
of which our solar system formed \cite{Lyons05}. To model these
processes, information on the electronic structure of all the CO
isotopologs, including those with $^{17}$O, is needed.

CO is an extremely stable molecule with a dissociation energy of 11.09
eV, corresponding to a threshold of 1118 \AA. It took until the late
1980s to establish that no continuum absorption occurs longward of 912
\AA, but that CO photodissociation is dominated by line absorptions,
most of which are strongly predissociated \cite{Letzelter87,Stark91}.
These early laboratory data were used to build a detailed model of the
CO photodissociation rate in an interstellar cloud
\cite{Viala88,vDB88}.  The depth dependence of the CO
photodissociation is affected not only by self-shielding, but also by
mutual shielding by H and H$_2$ because these species absorb in the
same wavelength region. $^{12}$CO, in turn, can shield the less
abundant $^{13}$CO, C$^{18}$O and C$^{17}$O species, with the amount
of shielding depending on the wavelength shifts in the absorption
spectrum.  Thus, a complete numerical simulation of the entire
spectrum of $^{12}$CO, its isotopologs, H$_2$ and H is required to
correctly compute the attenuation at each depth into a cloud
\cite{Lee96,LePetit06}.

In the 20 years since these first models, a steady stream of new
laboratory data has become available on this key process. In the late
1980s, many of the bands recorded in the laboratory were for $^{12}$CO
only and had no electronic or vibrational designations, so that
simulation of the isotopolog spectra involved much guesswork. In
particular, the magnitude of the isotopolog shifts depends sensitively
on whether $v'$ in the upper state is zero or not. Moreover,
predissociation rates were generally not known and were simply assumed
to be unity. High-resolution spectra for many important bands of
$^{12}$CO, $^{13}$CO, C$^{18}$O and $^{13}$C$^{18}$O were subsequently
measured with synchrotron light sources
\cite{Eidelsberg92,Eidelsberg04,Eidelsberg06,Stark91}. The fact that
the experimental oscillator strengths have now been reconciled with
values inferred from astronomical measurements leaves little doubt about
their accuracy \cite{Sheffer03}.  Ultra-high-resolution spectra of
selected states have been obtained with VUV lasers
\cite{Ubachs94}. These exquisite data provide not only positions down
to 0.003 cm$^{-1}$ accuracy, but also measure directly the line widths
and thus the predissociation probabilities. This is especially
important for the $C$~$^1\Sigma^+$ and $E$~$^1\Pi$ states responsible
for most of the isotope-selective effects
\cite{Cacciani95,Cacciani01,Ubachs00,Cacciani04}.

Visser et al.\ \cite{Visser09} used the recent laboratory data to
develop a new model for the CO isotope-selective photodissociation
including the $^{17}$O isotopologs, and applied it to interstellar
clouds and disks around young stars. They also computed shielding
functions for a much larger range of CO excitation temperatures and
Doppler broadenings.  Although the overall rate has changed by only
30\%, from $2.0$ to $2.6 \times 10^{-10}$ s$^{-1}$ for the standard
interstellar radiation field \cite{Draine78}, the modeled depth
dependence differs significantly from earlier work. In particular, the
isotope selective effects are diminished by a factor of three or more
at temperatures above 100 K.  Note, however, that even this new study
\cite{Visser09} still had to make many assumptions on line positions,
oscillator strengths and predissociation rates for the minor
isotopologs, since high-resolution spectra of these minor species have
not yet been measured or published. Unexpected differences of up to an
order of magnitude have been found between oscillator strengths of
$^{12}$CO and $^{13}$CO for the same bands, complicating
extrapolations from the main isotopologs. These differences likely
arise from mixing between various electronic states that depend
sensitively on the relative location of the energy levels and may thus
differ for a specific ro-vibronic $v',J'$ level of a particular
isotopologue, as found for the case of N$_2$ \cite{Stark08}.

Another set of experiments on the CO isotopolog photodissociation has
been carried out using the Berkeley synchrotron source
\cite{Chakraborty08} . In contrast with the Paris and Amsterdam
results, these data are at comparatively low spectral resolution and
were analyzed much more indirectly by following a set of subsequently
occurringchemical reactions. The data were interpreted to imply
different predissociation probabilities for individual levels of
isotopologs caused by near-resonance accidental predissociation
processes. In this interpretation, the isotope-selective effects found
in meteorites would not require self- or mutual shielding processes.
While accidental pre-dissociation is not excluded, these experiments
and their conclusions have been challenged on many grounds and are not
supported by the higher resolution data cited above
\cite{Visser09,Federman09,Lyons09,Yin09}. This episode demonstrates
that there is no substitute for high-quality molecular physics data in
which individual unsaturated lines are recorded, in order to draw
astrophysical conclusions.

Another very stable molecule for which the photodissociation occurs
primarily by line absorptions is N$_2$. Thanks to decades of
laboratory experiments, the line oscillator strengths and
predissociation probabilities of the excited ro-vibrational states are
now known with very high accuracy. The N$_2$ and $^{14}$N$^{15}$N data
have recently been summarized and applied to interstellar chemistry
\cite{Li13,Heays14}.

\begin{figure}[t]
\centerline{\hbox{
\includegraphics[width=0.7\textwidth,angle=0]{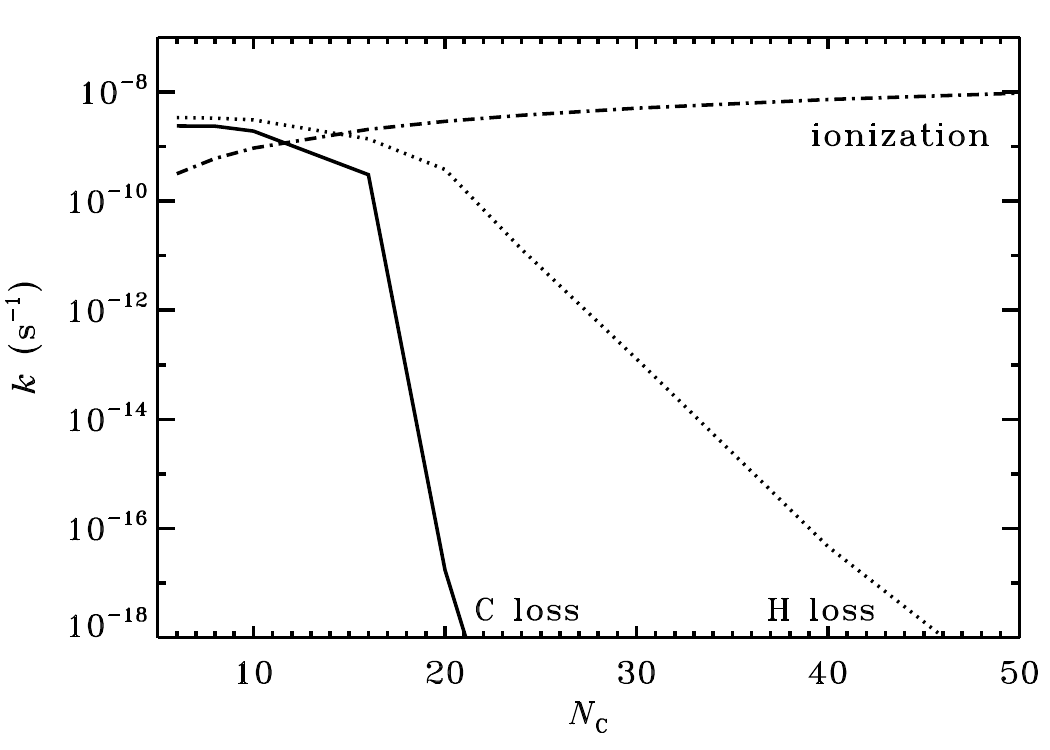}
}}
\caption{Photodissociation and photoionization rates of PAHs as
  functions of number of carbon atoms for the general interstellar
  radiation field. Solid: loss of C, dotted: loss of H; dash-dotted:
  ionization. Note the stability of the larger PAHs against
  dissociation. Figure made using data from \cite{Visser07}
  following \cite{Allain96}.}
\label{fig:PAHdis}
\end{figure}

\section{Photostability of PAHs}

PAH molecules are a class of large molecules that are very stable
against photodissociation in the general interstellar medium,
explaining their ubiquitous presence in galaxies
\cite{Tielens08}. However, in clouds exposed to very intense
radiation, such as the inner regions of protoplanetary disks and the
immediate environments of active galactic nuclei, the smaller PAHs can
be destroyed within the lifetime of the source. As discussed in Sec.\
4.2.2, the dissociation process of a large molecule involves a
multistep process, in which the molecule first absorbs a photon and is
promoted to an excited electronic state, then decays non-radiatively
to high vibrational levels of the ground state, and finally finds a
path to dissociation. The rate of this process thus depends not only
on the initial absorption rate, but also on the competition with other
processes during these steps. In the first step, the absorption of the
photon can also lead to emission of an electron through ionization or
photodetachment with probability $\eta_{\rm em}$. In the final step,
there is competition between dissociation and cooling by infrared
emission. The PAH photodissociation rate thus becomes
\cite{Leger89,Allain96,LePage01,Habart04,Visser07}:
$$ k_{\rm pd}= \int (1-\eta_{\rm em})\eta_{\rm diss} \sigma_{\rm abs} I_{\lambda} d\lambda $$
where $\eta_{\rm diss}$
is the yield of a single dissociation process given by
$$ \eta_{\rm diss}= {\sum_X {k_{\rm diss,X}}\over {(k_{\rm diss} + k_{\rm IR})}}. $$
Here $k_{\rm diss,X}$ is the rate for a particular loss channel $X$
(e.g., H removal), $k_{\rm diss}$ is the sum over all such channels,
and $k_{\rm IR}$ is the infrared emission rate. Possible loss channels
are H, H$_2$, C, C$_2$ and C$_3$ \cite{Leger89}, with the rates
determined by the RRKM quasi-equilibrium theory according to
$$ k_{{\rm diss,}X} = A_X {{\rho(E_{\rm int}-E_{0,X})} \over {\rho(E_{\rm int})}} $$
where $A_X$ is the pre-exponential Arrhenius factor for channel $X$
and $\rho(E)$ is the density of vibrational states at energy
$E$. Dissociation only occurs if the internal energy of the molecule
$E_{\rm int}$ exceeds the critical energy $E_0$ for a particular loss
channel.  Values for $E_{0,X}$ were summarized by
\cite{Visser07}.  The above formulation needs to be modified for
clouds exposed to very intense radiation fields ($G_0>10^4$), when
multiphoton events start to become significant, i.e., the PAH molecule
absorbs another UV photon before it has cooled down completely. This
can significantly increase the dissociation rate.

\begin{figure}
  \centerline{\hbox{
      \includegraphics[width=0.9\textwidth,angle=0]{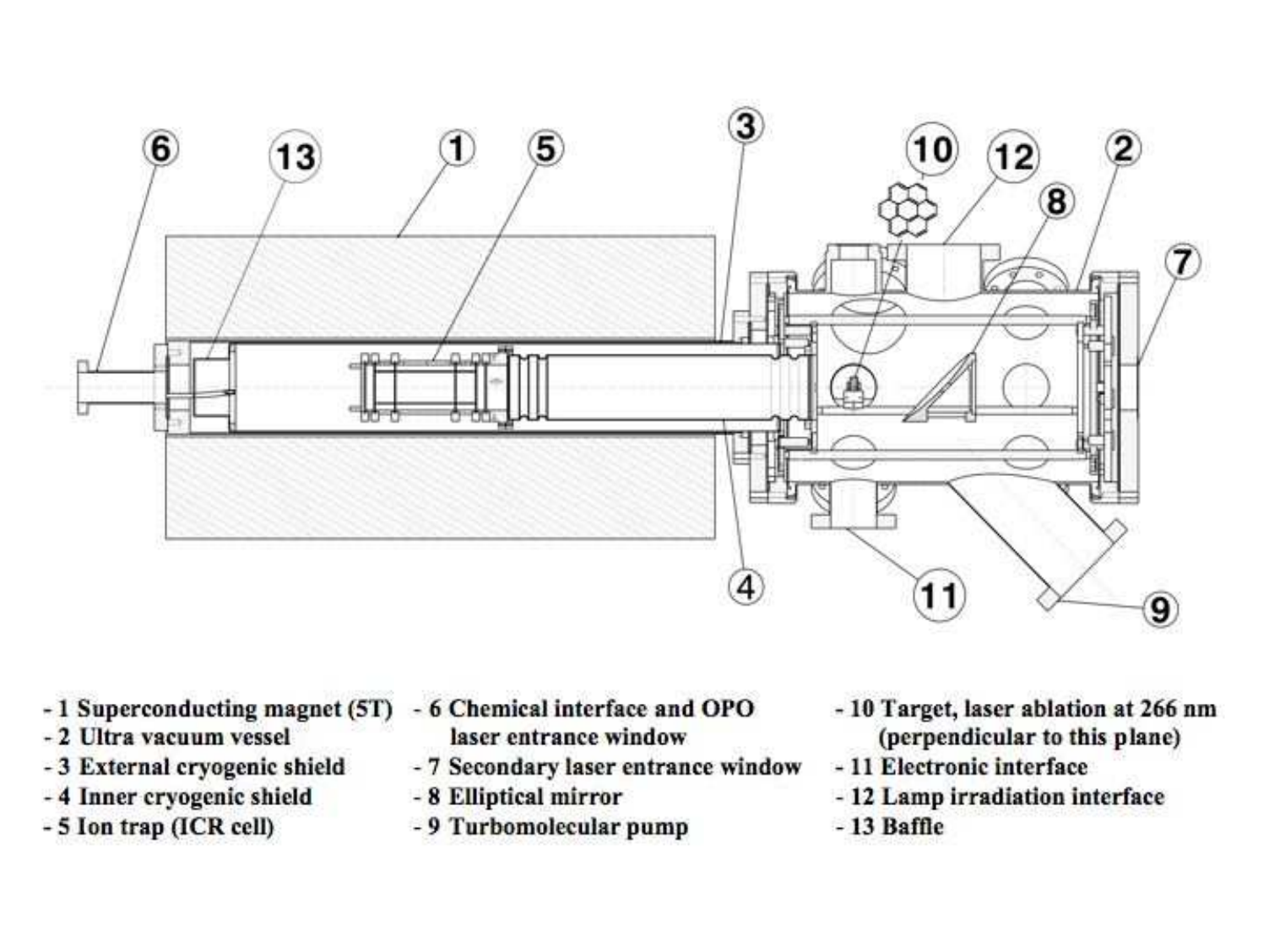}
    }}
  \caption{Overview of the PIRENEA experiment to study the
    photostability and fragmentation pattern of PAHs exposed to UV
    radiation \cite{Joblin02}.}
  \label{fig:pirenea}
\end{figure}

To compute the dissociation rates, values of $\sigma_{\rm abs}$ as a
function of wavelength are essential. The data used in astrophysical
models largely come from models developed by \cite{Li01}, based on
experiments by \cite{Joblin92}. There is also limited experimental
information on the loss channels, as summarized by
\cite{LePage01}. For example, \cite{Jochims94,Jochims99} have measured the
photostability of small PAHs of various sizes and shapes against
H-loss. Figure~\ref{fig:PAHdis} compares the H-loss channel
quantitatively with that of other loss channels and illustrates the
competition with ionization.

In a novel experiment, called PIRENEA, the spectroscopy and
photodissociation of PAHs into various fragments can be studied
\cite{Joblin02,Joblin11,Useli10}.  Specifically, gas-phase PAH
cations are produced by laser irradiation of a solid target and then
guided into an ion cyclotron resonance cell, where they are trapped in
a combined magnetic and electric field. The stored ions are irradiated
by a xenon lamp (2000--8000 \AA) and the products are analyzed by
Fourier-transform mass spectrometry. Ions of interest can be selected
and isolated by selective ejection of other species.  Initial results
show that irradiation leads predominantly to atomic hydrogen atom
loss, consistent with the work of \cite{Jochims94}, but loss of
C$_2$H$_2$ and perhaps even C$_4$H$_2$ is also seen from
dehydrogenated PAHs \cite{Useli07}.

\section{Summary}

This chapter has summarized our understanding of basic
photodissociation processes and the theoretical and experimental
approaches to determine cross sections, product branching ratios and
rates for astrophysically relevant species. The demand for accurate
photorates is likely to increase in the coming years as new infrared
and submillimeter facilities widen the study of molecules in the
surfaces of protoplanetary disks and exoplanetary
atmospheres. Critical evaluation of the photodissociation data for
these different environments are needed, since rates or cross sections
determined for interstellar clouds cannot be simply transposed to
other regions.

\section{Acknowledgments}

The authors are grateful to Marc van Hemert and Geert-Jan Kroes for
many fruitful collaborations on theoretical studies of
photodissociation processes, and to Roland Gredel, Christine Joblin
and Dave Parker for providing figures.  Support from a Spinoza grant
and grant 648.000.002 by the Netherlands Organization of Scientific
Research (NWO) and from the Netherlands Research School for Astronomy
(NOVA) is gratefully acknowledged.




\begin{thebibliography}{10}

\bibitem{vanD88}
E.~F. {van Dishoeck}.
\newblock {\em {Photodissociation and photoionization processes}}, volume 146
  of {\em Astrophysics and Space Science Library}, pages 49--72.
\newblock Springer, Berlin, 1988.

\bibitem{vanD06}
E.~F. {van Dishoeck}, B.~{Jonkheid}, and M.~C. {van Hemert}.
\newblock {\em Faraday Discussions}, 133:231, 2006.

\bibitem{Kirby88}
K.~P. {Kirby} and E.~F. {Van Dishoeck}.
\newblock {\em Advances in Atomic and Molecular Physics}, 25:437--476, 1989.

\bibitem{Schinke93}
R.~{Schinke}.
\newblock {\em {Photodissociation Dynamics}}.
\newblock Cambridge University Press, Cambridge, 1993.

\bibitem{vanD84}
E.F. van Dishoeck and A.~Dalgarno.
\newblock {\em Astrophysical Journal}, 277:576, 1984.
\newblock p. 576.

\bibitem{Chowdary08}
P.D. Chowdary and M.~Gruebele.
\newblock {\em Physical Review Letters}, 101:250603, 2008.
\newblock p. 250603.

\bibitem{Ashfold10}
M.N.R. Ashfold, G.A. King, and D.~et~al. Murdock.
\newblock {\em Phys. Chem. Chem. Phys.}, 12:1218--38, 2010.
\newblock p. 1218.

\bibitem{teVelde01}
G.~te~Velde, F.M. Bickelhaupt, S.J.A. van Gisbergen, et~al.
\newblock {\em Journal of Computational Chemistry}, 22:931--967, 2001.
\newblock p. 931 , www.scm.com.

\bibitem{Shao06}
Y.~{Shao}, L.~F. {Molnar}, Y.~{Jung}, et~al.
\newblock {\em Phys. Chem. Chem. Phys.}, 8:3172, 2006.
\newblock p. 3172.

\bibitem{Buenker74}
R.J. Buenker and S.D. Peyerimhoff.
\newblock {\em Theor. Chim. Acta}, 39:--, 1974.
\newblock p. 217.

\bibitem{Buenker78}
R.J. Buenker, S.D. Peyerimhoff, and W.~Butscher.
\newblock {\em Molecular Physics}, 35:771--791, 1978.
\newblock p. 771.

\bibitem{Werner10}
H.-J. et~al. Werner.
\newblock {\em MOLPRO, version 2010.1}, 1:0, 2010.
\newblock www.molpro.net.

\bibitem{Kroes93}
G.J. Kroes, E.F. van Dishoeck, R.~Be\"arda, and M.C. van Hemert.
\newblock {\em J. Chem. Phys.}, 99:228--236, 1993.
\newblock p. 228.

\bibitem{Bearda95}
R.A. Be\"arda, M.C. van Hemert, and E.F. van Dishoeck.
\newblock {\em The Journal of Chemical Physics}, 102:8930--8941, 1995.
\newblock p. 8930.

\bibitem{Fillion01}
JH~Fillion, R~van Harrevelt, J~Ruiz, et~al.
\newblock {\em {J. Phys. Chem. A}}, {105}({51}):{11414--11424}, {2001}.

\bibitem{vanHemert08}
M.C. van Hemert and E.F. van Dishoeck.
\newblock {\em Chem. Phys.}, 343:292--302, 2008.
\newblock p. 292.

\bibitem{Lee84}
L.C. Lee.
\newblock {\em Astrophys. J.}, 282:172--177, 1984.
\newblock p. 172.

\bibitem{Hudson71}
R.D. Hudson.
\newblock {\em Rev. Geophys. Space Phys.}, 9:305--406, 1971.
\newblock p. 305.

\bibitem{Okabe78}
H.~{Okabe}.
\newblock {\em {Photochemistry of Small Molecules}}.
\newblock Wiley, New York, 1978.

\bibitem{Ubachs94}
W.~Ubachs, K.S.E. Eikema, and P.F. Levelt.
\newblock {\em Astrophys. J.}, 728:L55--L58, 1994.
\newblock p. L55.

\bibitem{Pilling07}
S.~Pilling, R.~Neves, A.C.F. Santos, and H.M. Boechat-Roberty.
\newblock {\em Astron. Astrophys.}, 464:393--398, 2007.
\newblock p. 393.

\bibitem{Chandler87}
D.W. Chandler and P.L. Houston.
\newblock {\em J. Chem. Phys.}, 87:1445--1447, 1987.
\newblock p. 1445.

\bibitem{Eppink97}
Andre T. J.~B. Eppink and David~H. Parker.
\newblock {\em Review of Scientific Instruments}, 68(9), 1997.

\bibitem{Ashfold06}
M.~N.~R. {Ashfold}, N.~H. {Nahler}, A.~J. {Orr-Ewing}, et~al.
\newblock {\em Physical Chemistry Chemical Physics (Incorporating Faraday
  Transactions)}, 8:26, 2006.

\bibitem{Chabot10}
M.~{Chabot}, T.~{Tuna}, K.~{B{\'e}roff}, et~al.
\newblock {\em \aap}, 524:A39, 2010.

\bibitem{Storzer98}
H.~{Storzer} and D.~{Hollenbach}.
\newblock {\em \apjl}, 502:L71, 1998.

\bibitem{Acke06}
B.~{Acke} and M.~E. {van den Ancker}.
\newblock {\em \aap}, 449:267--279, 2006.

\bibitem{Oberg09}
K.~I. {{\"O}berg}, R.~T. {Garrod}, E.~F. {van Dishoeck}, and H.~{Linnartz}.
\newblock {\em \aap}, 504:891--913, 2009.

\bibitem{Black94}
J.H. Black.
\newblock {\em ASP conference series}, 58:--, 1994.
\newblock p. 355.

\bibitem{vanD94}
E.F. van Dishoeck.
\newblock {\em ASP Conf. Ser.}, 59:276, 1994.
\newblock p. 319.

\bibitem{Habing68}
H.~J. {Habing}.
\newblock {\em \bain}, 19:421, 1968.

\bibitem{Draine78}
B.~T. {Draine}.
\newblock {\em \apjs}, 36:595--619, 1978.

\bibitem{Parravano03}
A.~{Parravano}, D.~J. {Hollenbach}, and C.~F. {McKee}.
\newblock {\em \apj}, 584:797--817, 2003.

\bibitem{vDB82}
E.~F. {van Dishoeck} and J.~H. {Black}.
\newblock {\em \apj}, 258:533--547, 1982.

\bibitem{Hauschildt99}
P.~H. {Hauschildt}, F.~{Allard}, J.~{Ferguson}, E.~{Baron}, and D.~R.
  {Alexander}.
\newblock {\em \apj}, 525:871--880, 1999.

\bibitem{Henry80}
R.~C. {Henry}, R.~C. {Anderson}, and W.~G. {Fastie}.
\newblock {\em Astrophys. J.}, 239:859--866, 1980.

\bibitem{Kamp01}
I.~{Kamp} and G.-J. {van Zadelhoff}.
\newblock {\em \aap}, 373:641--656, 2001.

\bibitem{Bergin03}
E.~A. {Bergin}, M.~J. {Kaufman}, G.~J. {Melnick}, R.~L. {Snell}, and J.~E.
  {Howe}.
\newblock {\em \apj}, 582:830--845, 2003.

\bibitem{Hollenbach79}
D.~{Hollenbach} and C.~F. {McKee}.
\newblock {\em \apjs}, 41:555--592, 1979.

\bibitem{Neufeld89}
D.~A. {Neufeld} and A.~{Dalgarno}.
\newblock {\em \apj}, 340:869--893, 1989.

\bibitem{Gredel87}
R.~{Gredel}.
\newblock {\em {Fluorescent and collisional excitation in diatomic molecules.}}
\newblock Diss. Naturwiss.-Math. Gesamtfak., Ruprecht-Karls-Univ., Heidelberg,
  1987.

\bibitem{Gredel89}
R.~{Gredel}, S.~{Lepp}, A.~{Dalgarno}, and E.~{Herbst}.
\newblock {\em \apj}, 347:289--293, 1989.

\bibitem{Cardelli88}
J.~A. {Cardelli} and B.~D. {Savage}.
\newblock {\em \apj}, 325:864--879, 1988.

\bibitem{vDB89}
E.~F. {van Dishoeck} and J.~H. {Black}.
\newblock {\em \apj}, 340:273--297, 1989.

\bibitem{Roberge91}
W.~G. {Roberge}, D.~{Jones}, S.~{Lepp}, and A.~{Dalgarno}.
\newblock {\em \apjs}, 77:287--297, 1991.

\bibitem{Vasyunin11}
A.~I. {Vasyunin}, D.~S. {Wiebe}, T.~{Birnstiel}, et~al.
\newblock {\em \apj}, 727:76, 2011.

\bibitem{Dalgarno70}
A.~{Dalgarno} and T.~L. {Stephens}.
\newblock {\em \apjl}, 160:L107, 1970.

\bibitem{Werner70}
M.~W. {Werner}.
\newblock {\em \aplett}, 6:81, 1970.

\bibitem{Huebner92}
W.F. Huebner, J.J. Keady, and S.P. Lyon.
\newblock {\em Astrophys. Space Sci.}, 195, 1992.

\bibitem{Lyons05}
J.~R. {Lyons} and E.~D. {Young}.
\newblock {\em \nat}, 435:317--320, 2005.

\bibitem{Letzelter87}
C.~{Letzelter}, M.~{Eidelsberg}, F.~{Rostas}, J.~{Breton}, and
  B.~{Thieblemont}.
\newblock {\em Chemical Physics}, 114:273--288, 1987.

\bibitem{Stark91}
G.~{Stark}, K.~{Yoshino}, P.~L. {Smith}, K.~{Ito}, and W.~H. {Parkinson}.
\newblock {\em \apj}, 369:574--580, 1991.

\bibitem{Viala88}
Y.~P. {Viala}, C.~{Letzelter}, M.~{Eidelsberg}, and F.~{Rostas}.
\newblock {\em \aap}, 193:265--272, 1988.

\bibitem{vDB88}
E.~F. {van Dishoeck} and J.~H. {Black}.
\newblock {\em \apj}, 334:771--802, 1988.

\bibitem{Lee96}
H.-H. {Lee}, E.~{Herbst}, G.~{Pineau des Forets}, E.~{Roueff}, and J.~{Le
  Bourlot}.
\newblock {\em \aap}, 311:690--707, 1996.

\bibitem{LePetit06}
F.~{Le Petit}, C.~{Nehm{\'e}}, J.~{Le Bourlot}, and E.~{Roueff}.
\newblock {\em \apjs}, 164:506--529, 2006.

\bibitem{Eidelsberg92}
M.~{Eidelsberg}, J.~J. {Benayoun}, Y.~{Viala}, et~al.
\newblock {\em Astron. Astrophys.}, 265:839--842, 1992.

\bibitem{Eidelsberg04}
M.~{Eidelsberg}, F.~{Launay}, K.~{Ito}, et~al.
\newblock {\em J. Chem. Phys.}, 121:292--308, 2004.

\bibitem{Eidelsberg06}
M.~{Eidelsberg}, Y.~{Sheffer}, S.~R. {Federman}, et~al.
\newblock {\em \apj}, 647:1543--1548, 2006.

\bibitem{Sheffer03}
Y.~{Sheffer}, S.~R. {Federman}, and B.-G. {Andersson}.
\newblock {\em \apjl}, 597:L29--L32, 2003.

\bibitem{Cacciani95}
P.~{Cacciani}, W.~{Hogervorst}, and W.~{Ubachs}.
\newblock {\em J. Chem. Phys.}, 102:8308--8320, 1995.

\bibitem{Cacciani01}
P.~{Cacciani}, F.~{Brandi}, I.~{Velchev}, et~al.
\newblock {\em European Physical Journal D}, 15:47--56, 2001.

\bibitem{Ubachs00}
W.~Ubachs, I.~Velchev, and P.~Cacciani.
\newblock {\em The Journal of Chemical Physics}, 113(2), 2000.

\bibitem{Cacciani04}
P.~{Cacciani} and W.~{Ubachs}.
\newblock {\em Journal of Molecular Spectroscopy}, 225:62--65, 2004.

\bibitem{Visser09}
R.~{Visser}, E.~F. {van Dishoeck}, S.~D. {Doty}, and C.~P. {Dullemond}.
\newblock {\em \aap}, 495:881--897, 2009.

\bibitem{Stark08}
G.~{Stark}, B.~R. {Lewis}, A.~N. {Heays}, et~al.
\newblock {\em J. Chem. Phys.}, 128(11):114302, 2008.

\bibitem{Chakraborty08}
S.~{Chakraborty}, M.~{Ahmed}, T.~L. {Jackson}, and M.~H. {Thiemens}.
\newblock {\em Science}, 321:1328--, 2008.

\bibitem{Federman09}
S.~R. {Federman} and E.~D. {Young}.
\newblock {\em Science}, 324:1516--, 2009.

\bibitem{Lyons09}
J.~R. {Lyons}, E.~A. {Bergin}, F.~J. {Ciesla}, et~al.
\newblock {\em Geochimica et Cosmochimica Acta}, 73:4998--5017, 2009.

\bibitem{Yin09}
Qing-Zhu Yin, Xiaoyu Shi, Chao Chang, and Cheuk-Yiu Ng.
\newblock {\em Science}, 324(5934):1516, 2009.

\bibitem{Li13}
X.~{Li}, A.~N. {Heays}, R.~{Visser}, et~al.
\newblock {\em \aap}, 555:A14, 2013.

\bibitem{Heays14}
A.~N. {Heays}, R.~{Visser}, R.~{Gredel}, et~al.
\newblock {\em \aap}, page in press, 2014.

\bibitem{Visser07}
R.~{Visser}, V.~C. {Geers}, C.~P. {Dullemond}, et~al.
\newblock {\em \aap}, 466:229--241, 2007.

\bibitem{Allain96}
T.~{Allain}, S.~{Leach}, and E.~{Sedlmayr}.
\newblock {\em \aap}, 305:602, 1996.

\bibitem{Tielens08}
A.~G.~G.~M. {Tielens}.
\newblock {\em \araa}, 46:289--337, 2008.

\bibitem{Leger89}
A.~{Leger}, L.~{D'Hendecourt}, P.~{Boissel}, and F.~X. {Desert}.
\newblock {\em \aap}, 213:351--359, 1989.

\bibitem{LePage01}
V.~{Le Page}, T.~P. {Snow}, and V.~M. {Bierbaum}.
\newblock {\em \apjs}, 132:233--251, 2001.

\bibitem{Habart04}
E.~{Habart}, A.~{Natta}, and E.~{Kr{\"u}gel}.
\newblock {\em \aap}, 427:179--192, 2004.

\bibitem{Joblin02}
C.~{Joblin}, C.~{Pech}, M.~{Armengaud}, P.~{Frabel}, and P.~{Boissel}.
\newblock In M.~{Giard}, J.~P. {Bernard}, A.~{Klotz}, and I.~{Ristorcelli},
  editors, {\em EAS Publications Series}, volume~4 of {\em EAS Publications
  Series}, pages 73--77, 2002.

\bibitem{Li01}
A.~{Li} and B.~T. {Draine}.
\newblock {\em \apj}, 554:778--802, 2001.

\bibitem{Joblin92}
C.~{Joblin}, A.~{Leger}, and P.~{Martin}.
\newblock {\em \apjl}, 393:L79--L82, 1992.

\bibitem{Jochims94}
H.~W. {Jochims}, E.~{Ruhl}, H.~{Baumgartel}, S.~{Tobita}, and S.~{Leach}.
\newblock {\em \apj}, 420:307--317, 1994.

\bibitem{Jochims99}
H.~W. {Jochims}, H.~{Baumg{\"a}rtel}, and S.~{Leach}.
\newblock {\em \apj}, 512:500--510, 1999.

\bibitem{Joblin11}
C.~{Joblin} and A.~G.~G.~M. {Tielens}, editors.
\newblock {\em {PAHs and the Universe: A Symposium to Celebrate the 25th
  Anniversary of the PAH Hypothesis}}, volume~46 of {\em EAS Publications
  Series}, 2011.

\bibitem{Useli10}
F.~{Useli-Bacchitta}, A.~{Bonnamy}, G.~{Mulas}, et~al.
\newblock {\em Chemical Physics}, 371:16--23, 2010.

\bibitem{Useli07}
F.~{Useli Bacchitta} and C.~{Joblin}.
\newblock In {\em Molecules in Space and Laboratory}, 2007.

\end{thebibliography}
\end{document}